# Simulation Study of γγ -> hh in a Photon Collider

Tohru Takahashi[1], Nozomi Maeda[1], Katsumasa Ikematsu[2], Keisuke Fujii[2],
Eri Asakawa[3], Daisuke Harada[2], Shinya Kanemura[4], Yoshimasa Kurihara[2] and Yasuhiro Okada[2]

1 – Graduate School of Advanced Sciences of Matter – Hiroshima University
Higashi-Hiroshima, Hiroshima – Japan

2 – KEK, High Energy Accelerator Research Organization
Tsukuba, Ibaraki – Japan

3 – Institute of Physics,– Meiji Gakuin University
Yokohama, Kanagawa – Japan

4 – Department of Physics – University of Toyama
Toyama, Toyama – Japan

We studied a feasibility of measuring Higgs boson pair production in a Photon Linear Collider. The optimum energy of γγ collision was estimated with a realistic luminosity distribution. We also discussed simulation study for detecting the signal against W boson pair backgrounds.

## 1 Introduction

Discovery the Higgs boson, the missing part of the standard model, is urgent and important task for present particle physics and it is expected to be found at the LHC experiment. Assuming discovery of the Higgs boson, the precision measurement of its property, where the ILC electron-positron collider will play a main role, is a key to establish the mechanism of mass generation of gauge boson and fermions based on the spontaneous symmetry breaking. In addition, new physics beyond the standard model may manifest via the precision measurements of the Higgs sector.

Higgs physics at the ILC has been studied and summarized in [2]. In addition to the electron positron interaction at the ILC, an idea to construct the Photon Linear Collider (PLC) is considered as an option of the ILC. In the photon collider, intense laser pulses are flushed on to electrons accelerated by the linac and are converted to photon beams by the backward Compton scattering between laser photons and electrons. The PLC has potential to provide information of two photon decay with of the Higgs boson via $\gamma\gamma \to H$. The design of the PLC and physics opportunities has described in references [2-3].

Among Higgs boson properties, the mass and the self-coupling constant are to determine Higgs potential. While the measurement of the Higgs boson mass is the first thing to be done in the ILC experiment, the self-coupling constant requires detail investigation and luminosity to be measured. Thus, to see prospects to measure the self-coupling constant is important for the projection of the ILC performance to reveal nature of the electroweak symmetry breaking as well as to explore physics beyond the standard model which could manifest via symmetry breaking.

## 2 Pair Production of Higgs boson in the PLC

### 2.1 Higgs boson self-coupling in the PLC

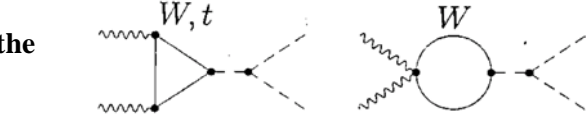

**Figure 1**: Diagrams for Higgs boson pair creation by γγ interaction via self-coupling

In the PLC, Higgs boson can be pair produced via self-coupling as shown in figure 1. It was first investigated in Ref. [4] and recently revisited [5]. The Higgs potential is expressed in general as;

$$V = \frac{1}{2} m_h^2 h^2 + \frac{1}{3!} \lambda_{hhh} h^3 + \frac{1}{4!} \lambda_{hhhh} h^4 + \cdots$$

where $m_h$ and $\lambda_{hhh}$, $\lambda_{hhhh}$ are parameters to be determined. Comparison with the Higgs potential in the standard model;

$$V = -\mu^2 |\phi|^2 + \lambda |\phi|^4$$

reads,
$m_h^2 = \lambda v^2$, $\lambda_{hhh} = \lambda v$, $\lambda_{hhhh} = \lambda$, respectively with $v$ being vacuum expectation value of the Higgs field. We define self-coupling parameter as;

$$\lambda = \lambda_{SM}(1 + \delta\kappa)$$

where $\lambda_{SM}$ and $\delta\kappa$ stand for Higgs self-coupling in the standard model and its deviation respectively.

As was discussed in [5], Higgs self-coupling can be studied in the electron positron interaction via $e^+e^- \to Zhh, hh\nu\nu$. Unlike the electron positron interaction, Higgs boson is produced by loop induced processes. Thus $\delta\kappa$ dependence is different from those in the electron positron interaction and the PLC has different sensitivity for new physics.

### 2.2 Optimized PLC Parameters

In order to set beam parameters of the PLC for Higgs self-coupling study by pair production. We first searched for the optimum beam energy to maximize statistical sensitivity, $S_{stat}$, which is defined as;

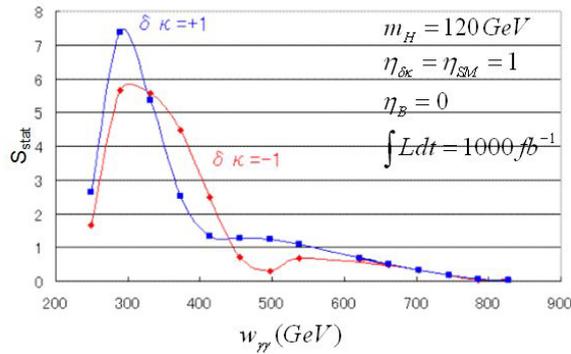

**Figure 2:** Sensitivity for Higgs boson self-coupling as a function of center of mass energy (CMS) of γγ collision. See detail for definition of the CMS energy

$$S_{stat} \equiv \frac{N(\delta\kappa) - N_{SM}}{\sqrt{N_{obs}}}$$

$$= \frac{L_{tot} \left| \eta_{\delta\kappa} \hat{\sigma}(\delta\kappa) - \eta_{SM} \sigma_{SM} \right|}{\sqrt{L_{tot} \left( \eta_{\delta\kappa} \hat{\sigma}(\delta\kappa) + \eta_B \sigma_B \right)}}.$$

$N(\delta\kappa), N_{SM}, N_{obs}$ are number of expected signal as a function of anomalous coupling parameter $\delta\kappa$, number of signal expected from the standard mode and number of observed events, respectively. $L_{tot}$ is total integrated luminosity defined as;

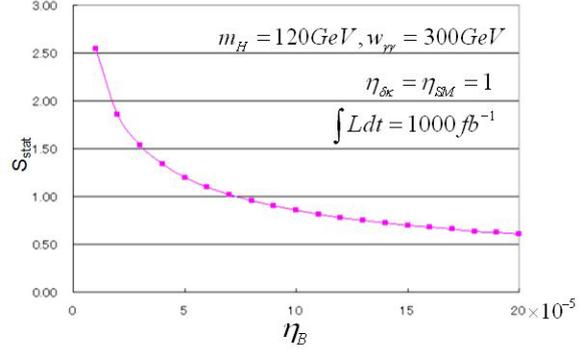

**Figure 3:** Sensitivity for anomalous coupling as a function of ratio of background (γγ->WW) suppression

$$L_{tot} \equiv \int \frac{dL}{dw_{\gamma\gamma}} dw_{\gamma\gamma}$$

while $\eta_i, \hat{\sigma}_j$ are efficiency and effective cross sections with B stands for background processes. The effective cross sections are estimated by convoluting theoretically calculated luminosity distribution as;

$$\hat{\sigma}_i \equiv \frac{1}{L_{tot}} \int \sigma_i(w_{\gamma\gamma}) \frac{dL}{dw_{\gamma\gamma}} dw_{\gamma\gamma}$$

The calculated sensitivity is shown in figure 2. In the calculation, Higgs boson mass of 120 GeV, $\eta_{\delta\kappa} = \eta_{SM} = 1$, $\eta_B = 0$ and $L_{tot}$ of $1000 fb^{-1}$ are assumed.
The horizontal axis of the figure is the high energy peak in the luminosity distribution in term of γγ collision energy [3]. The effective cross section for pair production $\hat{\sigma}_{SM}(\gamma\gamma \to hh)$ is typically 0.02fb. The main background process is W boson pair production and $\hat{\sigma}_B(\gamma\gamma \to W^+W^-)$ about 50 pb.
The study indicated that that the PLC parameters have to be set to realize luminosity peak to be around 300 GeV. Figure 3 shows the sensitivity as a function of the efficiency, $\eta_B$, for the background processes, $\hat{\sigma}_B(\gamma\gamma \to W^+W^-)$. We see that the background have to be suppressed to the level of $10^{-6}$ to reach $S_{stat} \sim 3$. The same analysis is performed with realistic luminosity distribution described next section and is found to be consistent each other.

## 2.3 Beam Parameters

**Table 1:** Parameters for the electron beam and the laser pulse. Refer text for the definition of parameters not stated in the table

| | | |
|---|---|---|
| Electron beam energy [GeV] | 210 | 195 |
| # of electron in a bunch ($10^{10}$) | 2 | 2 |
| Bunch length(mm) | 0.35 | 0.35 |
| Normalized emittance [m rad] | 2.5/0.03 | 2.5/0.03 |
| Beat function at the IP[mm] | 1.5/0.3 | 1.5/0.3 |
| Electron beam size at the IP[nm] | 96/4.7 | 99/5.5 |
| Laser Wave length[nm] | 1054 | 770 |
| Laser Pulse Energy[J] | 10 | 10 |
| x parameter | 3.76 | 4.8 |
| Geometric luminosity [$10^{34}$cm$^{-2}$s$^{-1}$] | 8.7 | 8.1 |
| Total $\gamma\gamma$ luminosity[$10^{34}$cm$^{-2}$s$^{-1}$] | 12.6 | 5.88 |
| Effective cross section[fb] | 0.0131 | 0.0205 |

Based on the discussion in the previous section, a set of parameters for the PLC were prepared as summarized in table 1. The maximum energy and energy distribution of the Compton scattered photons depend on a kinematical parameter

$$x \equiv \frac{4\omega E}{m^2 c^4}$$

where $\omega, E, m, c$ are energy of laser photon, electron beam energy, electron rest mass and velocity of light, respectively. Two parameter set were prepared for different x parameter taking technical issues in to account for the peak $\gamma\gamma$ center of mass energy of 300 GeV. A parameter with x=4.8 is typically chosen to maximize energy of the PLC while the one with x=3.76 is to keep wave length of the laser at 1054nm which is a technically preferable for solid state lasers. Figure 4 shows luminosity distribution with x=4.8 simulated with the CAIN program[6]. The integrated luminosity, effective cross section are also summarized in table 1.

## 3 Simulation Study of Higgs pair production

### 3.1 Simulation Scheme

In order to estimate efficiency for signals and backgrounds, we performed simulation study. The cross section were calculated using program developed by authors and cross section of background were calculated using GRACE, an automatic event generator program [7]. During the calculation, luminosity distribution described in section 2.3 was convoluted. Finally, detector response were simulated by the JSF quick simulator [8].

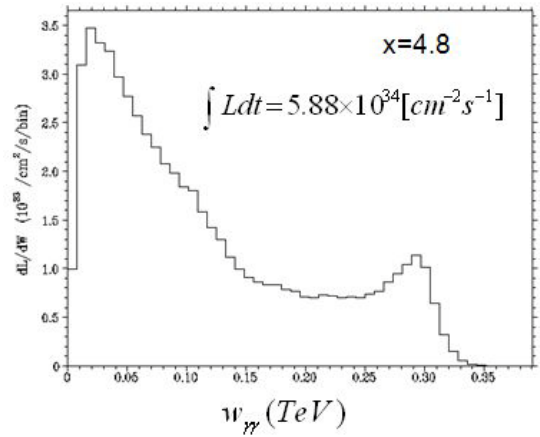

**Figure 4:** Simulated luminosity distribution by CAIN with x=4.8 parameters

## 3.2 Event Analysis

Since Higgs boson of 120 GeV mainly decays into b-quark pairs with the branching ratio of about 0.67, we concentrated to the case that both Higgs boson decayed into b-quark pairs. For each event from the JSF simulator, we applied so called forced four jets analysis in which the clustering algorithm is applied to the event by changing clustering parameter until the event is categorized as a four jets event. After the forced four jets analysis, invariant masses for jet pairs were calculated. As it is a four jets event, it is necessary to choose a right jets pairs originating from parent Higgs (W for the background) bosons out of three possible combinations.

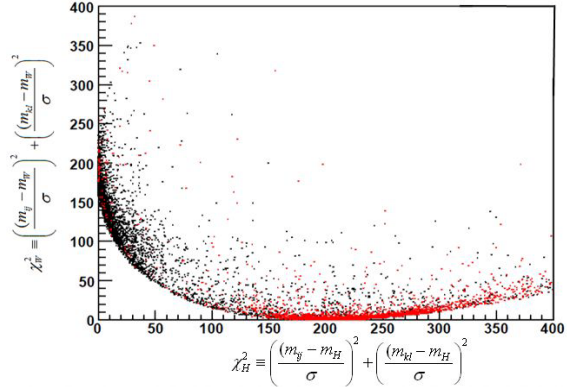

**Figure 5:** $\chi^2$ for two mass assumptions. One combination giving least $\chi^2$ for Higgs or W boson mass assumption was chosen for an event. The black points are for Higgs event and reds are for W bosons

For this purpose we defined a $\chi^2$ as;

$$\chi^2_{h(W)} \equiv \sum_i \frac{\left(m^i_{jj} - m_{h(W)}\right)^2}{\sigma^2},$$

where suffix h(w) means the $\chi^2_{h(W)}$ assuming Higgs(W) boson mass for a jet pairs. $\sigma$ is the mass resolution for a jet pairs and is estimated to be 4 GeV in average. The summation runs over two jet pairs for one combination. In each jet combination, $\chi^2$ was calculated for both Higgs and W mass assumption so that so that total of six $\chi^2$ s are calculated for an event. The jet combination and hypnoses of the least $\chi^2$ was chosen to be a most probable assumption for an event. Figure 4 show correlation of $\chi^2_h$ and $\chi^2_W$ for the most probable assumption in an event. It appears to be possible to separate signal and background by applying proper cut in the two dimensional plot.

## 4  Summary and Outlook

Aiming measurement of Higgs boson self-coupling in a PLC, feasibility study for detecting Higgs boson pairs production is being studied. The optimum center of mass energy of $\gamma\gamma$ for 120GeV Higgs boson was searched with both theoretical and with realistic luminosity distribution We found that both estimate consistently showed the optimum energy of about 300 GeV. Using parameter set for the electron beam and laser pulses, the number of events was estimated by convoluting luminosity distribution with the analytical calculation of the production cross section.

To investigate experimental feasibility, a trial to separate signal and W boson pair production background were performed. We found that the background can be suppressed once we apply cuts for kinematical variable such as properly defined $\chi^2$ using jet pair masses. However, event selection only with kinematical parameters did not appear to be enough to reach tolerable background contamination. It is concluded that development of sophisticated b-quark tagging for four jets event is crucial for further improvement.

## 5 Acknowledgements

The authors would like to thank the ILC physics working group for valuable discussion and suggestion.